# Emergent dynamical Kondo coherence and competing magnetic order in a correlated kagome flat-band metal CsCr$_6$Sb$_6$

Xiangqi Liu,[1, †] Xuefeng Zhang,[1, †] Jiachen Jiao,[2, †] Renjie Zhang,[3,4,5 †] Kaiwen Chen,[2] Ying Wang,[2] Yunguan Ye,[1] Zhenhai Yu,[1] Chengyu Jiang,[2] Xia Wang,[1,6] Lei Shu,[2,7, *] Baiqing Lv,[3, *] Gang Li,[1,8, *] and Yanfeng Guo[1,8, *]

[1]State Key Laboratory of Quantum Functional Materials, School of Physical Science and Technology, ShanghaiTech University, Shanghai 201210, China

[2]State Key Laboratory of Surface Physics, Department of Physics, Fudan University, Shanghai 200438, China

[3]Tsung-Dao Lee Institute, Shanghai Jiao Tong University, Shanghai 200240, China

[4]Beijing National Laboratory for Condensed Matter Physics and Institute of Physics, Chinese Academy of Sciences, Beijing 100190, China

[5]University of Chinese Academy of Sciences, Beijing 100049, China

[6]Analytical Instrumentation Center, School of Physical Science and Technology, ShanghaiTech University, Shanghai 201210, China

[7]Shanghai Research Center for Quantum Sciences, Shanghai 201315, China

[8]ShanghaiTech Laboratory for Topological Physics, ShanghaiTech University, Shanghai 201210, China

**Correlated kagome metals host unique electronic states that enable exotic quantum phenomena. In the recently emerged CsCr$_6$Sb$_6$, these manifest through Kondo behavior from localized Cr-3$d$ electrons and unprecedented band flattening near the Fermi level. Yet the intricate interplay among Kondo screening, magnetic frustration, and electronic correlations remains poorly understood—a fundamental gap we address through multifaceted experimental and theoretical approaches. Our angle-resolved photoemission spectroscopy measurements reveal**

**electronic correlation-renormalized flat bands and muon spin relaxation study detect short-range magnetic order at $T_N$ ~ 80 K. Complementing these findings, density-functional theory and dynamical mean-field theory calculations identify a coherent-incoherent crossover at $T_N$, with a remarkable restoration of coherence accompanying local moment suppression—an anomalous hallmark of Kondo behavior. Intriguingly, despite strong interlayer antiferromagnetic coupling, the system evades long-range magnetic order due to competing magnetic configurations separated by sub-meV energy differences. These insights establish $CsCr_6Sb_6$ as a prototypical platform for investigating dynamical Kondo screening in correlated flat-band systems, opening new avenues to study flat band physics and frustrated magnetism in correlated kagome lattices.**

[†]These authors contributed equally to this work.
Xiangqi Liu, Xuefeng Zhang, Jiachen Jiao, and Renjie Zhang.

[*]Correspondence:

[*]leishu@fudan.edu.cn,

[*]baiqing@sjtu.edu.cn,

[*]ligang@shanghaitech.edu.cn,

[*]guoyf@shanghaitech.edu.cn.

**Introduction**

Kagome lattices have emerged as a quintessential platform for discovering novel correlated topological states [1-22]. In their Mott-insulating phase, geometrically frustrated $S = 1/2$ moments give rise to quantum spin liquids featuring fractionalized excitations and emergent gauge fields [8,9]. When metallized, unique geometry of the lattice induces destructive interference in nearest-neighbor hopping, producing a remarkable electronic spectrum characterized by Dirac nodes, van Hove singularities (vHSs), and correlation-narrowed flat bands. These distinctive features, coupled with nontrivial Berry curvature, place kagome metals at the frontier where strong correlations meet topological physics, opening new avenues for discovering exotic quantum phases.

The $A$V$_3$Sb$_5$ compounds ($A$ = K, Rb, Cs) exemplify conventional kagome metals, featuring delocalized electrons from nonmagnetic V ions that form quasi-two-dimensional (2D) band structures tunable near vHS [1-3]. A fundamentally different scenario emerges when localized 4$f$/5$f$ electrons prevail, enhancing on-site Coulomb repulsion ($U$) and giving rise to Kondo-lattice physics — where conduction electrons screen local moments below a characteristic temperature scale [23-25]. While historically associated with 4$f$/5$f$ systems, Kondo phenomena have recently been identified in 3$d$ transition-metal oxides and van der Waals materials [26-30], prompting a reevaluation of the essential criteria for Kondo screening.

The manifestation of electronic correlations in kagome systems reveals remarkable diversity: Nb$_3$Cl$_8$ transforms into a Mott insulator with well-defined Hubbard bands [31,32], while Ni$_3$In and pressurized CsCr$_3$Sb$_5$ display strange-metal behavior mediated by flat-band Kondo effects without local moments [33-36]. The incipient flat band of CsCr$_3$Sb$_5$ (~ 80 meV below the Fermi level ($E_F$)) supports Hund's metallicity and antiferromagnetic (AFM) fluctuations, contrasting sharply with the Kondo insulating gap observed in the newly discovered CsCr$_6$Sb$_6$ [37]. The latter system provides unprecedented insight into the competition between Kondo screening and short-range magnetic order. Its ultra-flat band (band width ~ 40 meV) pinned at $E_F$ causes extreme electron localization, while spectroscopic evidence reveals a 10-15 meV Kondo gap — a hallmark of coherent moment screening. Most notably, the absence of long-range order despite strong local correlations suggests a quantum-critical regime where short-range magnetism and Kondo hybridization remain in delicate balance. This makes $CsCr_6Sb_6$ an exceptional platform for investigating: (1) Kondo coherence mechanisms in frustrated flat-band systems; (2) quantum fluctuation-mediated suppression of magnetic order in 3$d$-electron materials; and (3) short-range correlation effects on non-Fermi-liquid transport — key characteristics of strongly correlated quantum matter.

Our comprehensive study of $CsCr_6Sb_6$ combines experimental and theoretical approaches. Magnetotransport and magnetization measurements demonstrate characteristic heavy-fermion behavior with pronounced Kondo scattering at low temperatures. Angle-resolved photoemission spectroscopy (ARPES) reveals an isolated, dispersionless flat band precisely at $E_F$, creating an ideal platform for studying correlated magnetism in frustrated geometries. Unlike conventional Kondo lattices that develop coherent Fermi-liquid states through complete screening, zero-field muon spin relaxation (ZF-$\mu$SR) measurements detect only short-range magnetic order in $CsCr_6Sb_6$ — clear evidence of screening disrupted by persistent spin correlations. Density-functional theory (DFT) and dynamical mean-field theory (DMFT) calculations attribute this behavior to strong correlation effects that simultaneously renormalize the flat band, enhancing quasiparticle effective mass while preserving short-range spin correlations.

The details for single crystal growth, quality examinations, energy dispersive spectroscopy (EDS) characterizations, magneto-transport, ARPES and ZF-$\mu$SR measurements, and high-pressure modulation of the Kondo behavior in $CsCr_6Sb_6$ are presented in the Supporting Information (SI) which includes Figs. S1-S7 and Refs. [13, 38-52].

**Results and Discussions**

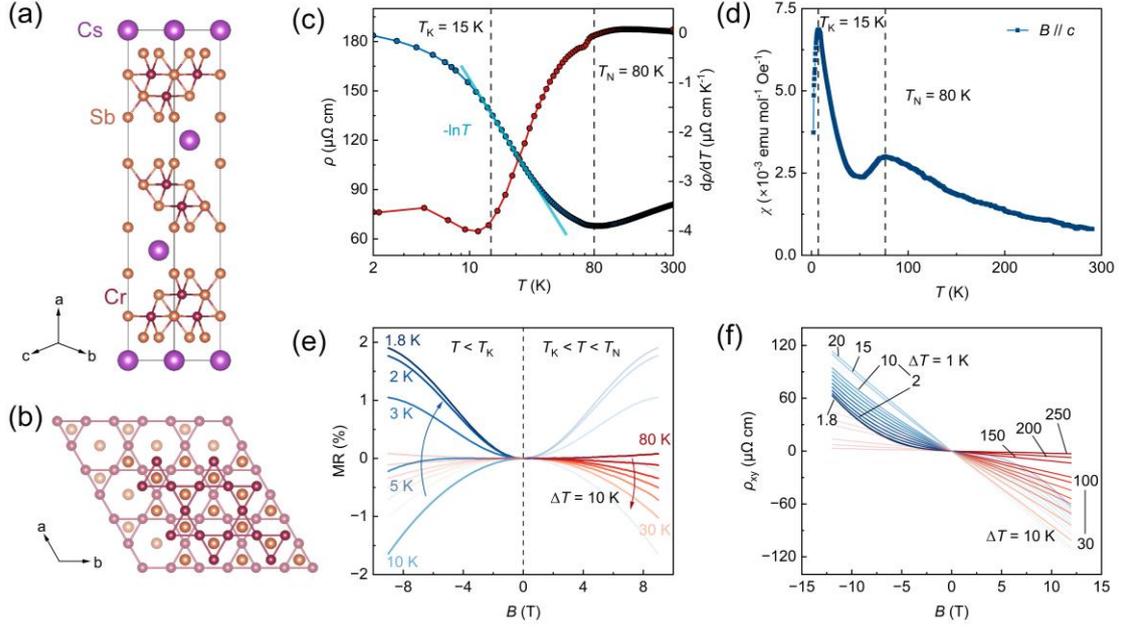

**Fig. 1.** (a) Side view of the crystal structure of $CsCr_6Sb_6$ showing the layered arrangement along the *c*-axis. (b) Top view of the double kagome network in the *ab*-plane, highlighting the geometrically frustrated Cr sublattice. (c) In-plane $\rho(T)$ (blue) and its derivative $d\rho/dT$ (red). The sharp change in $d\rho/dT$ at $T_N = 80$ K marks the onset of short-range magnetic order, below which enhanced Kondo scattering dominates. The cyan curve shows the characteristic $-\ln T$ dependence expected for Kondo scattering, with deviation below $T_K = 15$ K indicating partial formation of Kondo singlets. (d) $\chi(T)$ measured at $H = 2$ kOe ($H//c$). The magnetic susceptibility signal reveals: (i) Curie-Weiss behavior above $T_N$, (ii) enhanced spin correlations below $T_N$ and (iii) suppressed magnetism after the formalization of Kondo singlet. (e) MR and (f) Hall resistivity at selected temperatures ($H//c$ up to 12 T). The MR crossover from positive to negative below $T_N$ signifies suppressed spin-disorder scattering under magnetic fields. The negative MR peaks near $T_K$ before reverting to positive at lower temperatures, reflecting the competition between Kondo screening and field-induced spin polarization. Below $T_K$, the nonlinear Hall response reveals multiband behavior post-renormalization, while its strong temperature dependence underscores carrier density tuning by electronic correlations.

In contrast to $CsCr_3Sb_5$, which crystallizes in a hexagonal structure (space group: *P6/mmm*) with alternating 2D kagome $Cr_3Sb_5$ planes and Cs layers, $CsCr_6Sb_6$ adopts a rhombohedral lattice (space group: *R-3m*) featuring a double-layer kagome network ($Cr_6Sb_6$) stacked in an ABC sequence and intercalated by Cs spacers. As shown in Figs. 1(a,b), the V-Sb layers exhibit a lateral shift of (1/3*a*, -1/3*b*) along the *c*-axis, generating a geometrically frustrated lattice. This architecture hosts correlated 3*d* electrons that drive the observed Kondo phenomenology. The temperature-dependent resistivity $\rho(T)$ (Fig. 1(c)) reveals a metallic baseline ($d\rho/dT > 0$) interrupted by two anomalies: (i) a resistivity minimum at $T_N$ = 80 K, signaling the onset of short-range magnetic order, and (ii) a pronounced low-temperature upturn until $T_K$ (~ 15 K) indicative of Kondo scattering. This upturn is mirrored in the magnetic susceptibility $\chi(T)$ (Fig. 1(d)), which transitions from Curie-Weiss behavior at high temperatures to enhanced spin correlations below $T_N$. The low-temperature resistivity is quantitatively described by the phenomenological model: $\rho(T) = aT^5 + c\rho_0 - c\rho_1 \ln T$ [26], where $aT^5$ denotes unusual phonon resistivity, and the $-\ln T$ term reflects spin-flip scattering between localized Cr-3*d* moments and itinerant electrons. Below the Kondo temperature $T_K$ ~ 15 K, this logarithmic divergence softens as partial Kondo singlet formation reduces scattering—a hallmark of heavy-fermion behavior [27]. Parallels emerge with the *d*-electron Kondo system $Ni_3In$ [33], where frustration and flat bands similarly promote localized moments. In $CsCr_6Sb_6$, the flat band of kagome lattice (pinned near the $E_F$ by strong correlations) localizes Cr-3*d* electrons, creating a platform to explore Kondo physics without *f*-electron involvement.

The interplay of Kondo and magnetic order is further evidenced in magnetoresistance (MR, Fig. 1(e)). For $T > T_N$, a weak positive MR arises from the enhanced effective lattice scattering due to the Lorentz force. Below $T_N$, correlations drive local moments into short-range order, and a crossover to negative MR emerges as the field aligns the local moment—the scattering mechanism responsible for the resistivity enhancement, the amplitude of negative MR reaches a maximum at $T_K$. When the temperature decreases below $T_K$, a second crossover from negative to positive MR appears, the formation of Kondo singlets diminishes the local moment and screens the conducting electrons from scattering. At $T \ll T_K$, the increase in high-field MR becomes gradual once the Zeeman energy exceeds $k_B T_K$ ($k_B$ is the Boltzmann constant), consistent with the magnetic destruction of the Kondo singlets [28]. As evidenced in Fig. 1(f), the Hall response displays linear field dependence above $T_K$, with the negative Hall coefficient $R_H$ signifying dominant electron-type carriers in a single-band regime. Upon cooling below $T_K$, electronic correlations induce significant band renormalization, driving the system into a multiband phase. This transition manifests as pronounced nonlinearity in the Hall signals, which are quantitatively described by a two-carrier model (see SI, Section B for detailed analysis). The Hall effect measurements reveal a carrier density $n \sim 10^{19}$ cm$^{-3}$ at low temperatures, two orders of magnitude lower than in V-based kagome metals (e.g., CsV$_6$Sb$_6$, $n \sim 10^{21}$ cm$^{-3}$ [29]). This depressed $n$, comparable to low-carrier-density Kondo lattice Pr$_2$Ir$_2$O$_7$ [30], underscores the dominance of localized 3$d$ moments in charge transport. The correlation-driven band renormalization has been verified through experiments and calculations in Fig. 2 and Fig. 4, respectively. The above characterizations indicate that CsCr$_6$Sb$_6$ epitomizes a $d$-electron Kondo lattice where geometric frustration, flat-band localization, and strong correlations converge. The coexistence of short-range order ($T_N$ = 80 K) and Kondo screening ($T_K$ = 15 K) manifests in distinct transport and magnetic anomalies, offering a novel platform to explore correlated topology beyond traditional $f$-electron systems.

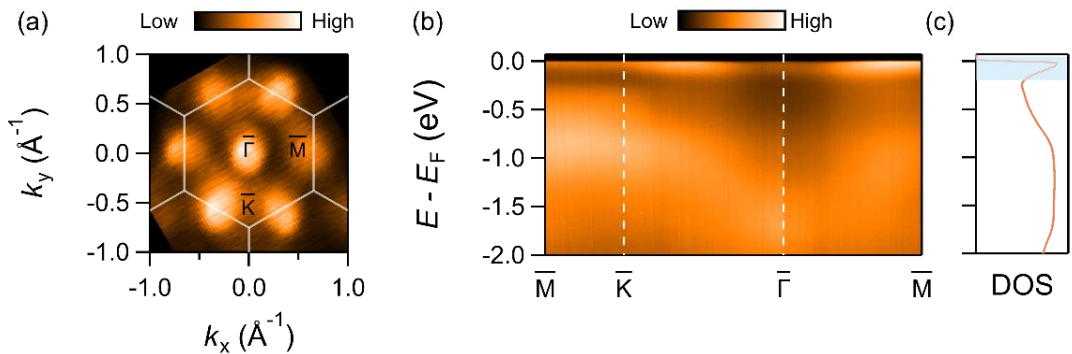

**Fig. 2.** (a) Fermi surface mapping of CsCr$_6$Sb$_6$ in the surface Brillouin zone, measured with $hv$ = 100 eV (linear horizontal polarization). (b) Band dispersion along the $\overline{M}$-$\overline{K}$-$\overline{\Gamma}$-$\overline{M}$ high-symmetry path ($hv$ = 30 eV, linear horizontal polarization). (c) Momentum-integrated DOS derived from (b).

The shaded region highlights the flat band's dominant contribution.

To elucidate the origin of $CsCr_6Sb_6$'s exotic properties, we performed high-resolution ARPES measurements. Fig. 2(a) presents the Fermi surface mapping in the surface Brillouin zone (perpendicular to the *c*-axis), revealing multiple electron pockets centered at high-symmetry points including two distinct circular pockets at the $\bar{\Gamma}$ and $\bar{M}$ points. A faint yet intriguing feature between $\bar{\Gamma}$ and $\bar{K}$ points, whose exact topology (a large pocket surrounding $\bar{\Gamma}$ versus a triangular pocket encircling $\bar{K}$) remains ambiguously. These features emerge from the intersection of multiple dispersive bands with the $E_F$, suggesting complex multiband physics at play. The band structure along the $\bar{M}$-$\bar{K}$-$\bar{\Gamma}$-$\bar{M}$ high-symmetry path (measured at $hv$ = 30 eV, Fig. 2(b)) unveils two key components, those are, a weakly dispersive band contributing to the observed Fermi surface pockets and a striking, non-dispersive flat band pinned near $E_F$. The flat band's exceptional uniformity is highlighted in the density of states (DOS, Fig. 2(b) inset), derived from spectral intensity integration. The DOS exhibits a sharp peak centered at -20 meV and a remarkably narrow energy spread (~ 200 meV). The peak's finite width likely reflects self-energy broadening due to strong electron correlations, which is a smoking gun for correlation-driven renormalization. This observation positions $CsCr_6Sb_6$ as a rare *d*-electron system where flat-band physics intertwines with substantial many-body effects.

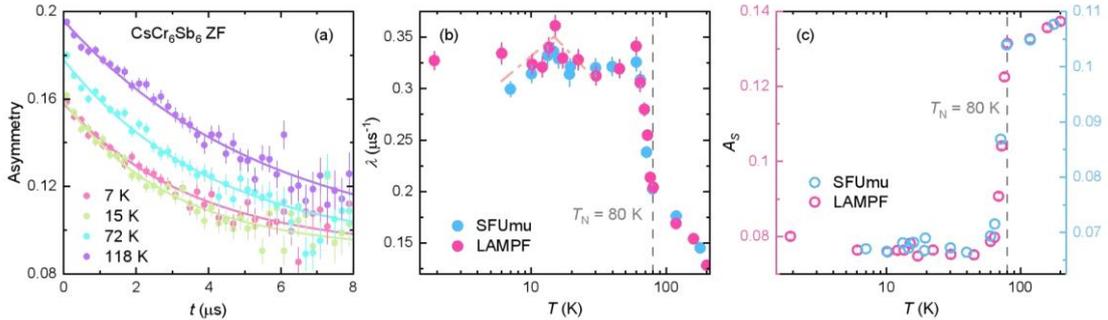

**Fig. 3.** (a) ZF-$\mu$SR spectra for $CsCr_6Sb_6$ at representative temperatures. Solid lines: fits to Eq. (1). (b)-(c) Temperature dependence of the relaxation rate $\lambda$ and $A_S$, respectively. Blue and Red circle: Data from M9

and M20 beamline, respectively.

The ZF-$\mu$SR spectra (Fig. 3(a)) collected from 2-300 K show no oscillatory components, demonstrating the absence of long-range magnetic order down to the base temperature. The spectra consist of two distinct contributions including an exponentially relaxing sample signal (characterized by initial asymmetry $A_S$ and relaxation rate $\lambda$) and a temperature-independent background signal ($A_{bg}$) originating from muons stopped in the silver sample holder. These features are well described by the function:

$$A_{sy}(t) = A_S \exp(-\lambda t) + A_{bg}. \tag{1}$$

Temperature-dependent analysis reveals a sharp transition in the initial asymmetry $A_S$ at $T_N \sim 80$ K (with a transition width of ~15 K), where its rapid suppression below $T_N$ indicates the formation of short-range magnetic order with an estimated volume fraction of 45%. This suppression reflects enhanced magnetic field inhomogeneity that causes muon spin depolarization faster than the instrumental time resolution (~ $10^{-7}$ s). Concurrently, the relaxation rate $\lambda$ exhibits a dramatic increase below $T_N$, signifying strengthened local magnetism. Notably, $\lambda$ displays a peak at ~ 15 K, coinciding with the Kondo temperature $T_K$ identified in transport and magnetic susceptibility measurements. These observations collectively demonstrate the coexistence of short-range magnetic order and Kondo screening in CsCr$_6$Sb$_6$, where the competition between these phenomena manifests in the distinct temperature evolution of both $A_S$ and $\lambda$. The $\mu$SR results provide compelling evidence for a complex magnetic ground state characterized by nanoscale order.

To elucidate the magnetic and electronic properties of CsCr$_6$Sb$_6$, we complemented our experimental investigations with theoretical calculations. ARPES measurements reveal a strongly renormalized flat band near $E_F$, which is absent in DFT band structures, suggesting a pivotal role of Coulomb interactions in band renormalization. To account for these strong electronic correlations, we conducted

DFT+DMFT calculations. Figs. 4(a)-(d) depict the DMFT-derived electronic spectral functions under varying $U$ and temperatures. For a modest $U = 2$ eV (Figs. 4(a)-(b)), the interacting spectra closely resemble the DFT bands. However, at a higher $U = 4$ eV (Fig. 4(c)), single-particle coherence is markedly suppressed at $T = 116$ K—a temperature exceeding the onset of Kondo screening. Notably, in addition to this loss of coherence, the band initially located at -0.4 to -0.2 eV below $E_F$ (Fig. 4(a)) shifts prominently toward $E_F$, while the dispersive band within -0.15 to 0.05 eV undergoes strong renormalization, forming a flat band at $E_F$ in excellent agreement with our ARPES data.

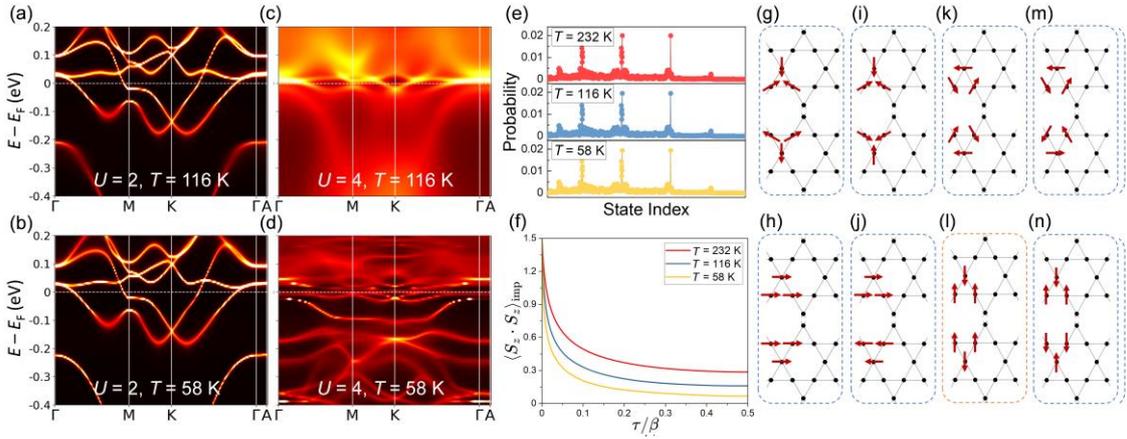

**Fig. 4.** (a)-(d) Interacting spectral functions as obtained by DFT + DMFT with $U = 2$, 4 eV and $T = 58$, 116 K. (e) Contribution probability of the atomic states and (f) Local spin-spin susceptibilities $\langle S_z(\tau) S_z(0) \rangle$ at $T = 58$, 116 and 232 K. (g)-(n) Competing magnetic configurations with in-plane moments examined in DFT total energy calculations. Each upper and lower kagome layer denote the two layers within a double kagome layer unit.

Intriguingly, as illustrated in Fig. 4(d), the single-particle coherence is restored at temperatures below $T_N$. This temperature-driven crossover from incoherent to coherent excitations bears resemblance to systems with long-range magnetic order. Yet, no such order exists in $CsCr_6Sb_6$. Instead, we propose that this crossover unambiguously signals enhanced Kondo screening at lower temperatures—a conclusion firmly supported by experimental evidence.

In conventional magnetically ordered systems, local moments exhibit strong fluctuations above the ordering temperature, resulting in the loss of single-particle excitation coherence. As the system cools below the ordering temperature, the gradual freezing of magnetic fluctuations leads to the restoration of coherent single-particle excitations, typically accompanied by an increase in local moment amplitude. Strikingly, $CsCr_6Sb_6$ presents a fundamentally different scenario—not only does it lack long-range magnetic order, but also our measurements reveal a progressive suppression of local moment amplitude with decreasing temperature. This unconventional behavior is quantitatively demonstrated through our DMFT analysis. Figs. 4(e) and 4(f) present the probabilities of DMFT atomic states and imaginary-time local spin-spin correlation functions, respectively, which provide direct insight into the temperature evolution of local moments. Fig. 4(e) reveals that the atomic states contributing to local moments remain essentially unchanged across the characteristic temperature $T_K$. The local susceptibilities in Fig. 4(f) track how conduction electrons dynamically screen the bare local moments in Cr-$3d$ orbitals. These correlations exhibit a rapid initial decay followed by slow saturation to small finite values—a clear signature of strong dynamical screening. Most remarkably, as temperature decreases, the long-time limit of $\langle S_z(\tau = \beta/2)S_z(0)\rangle$ shows further suppression, unambiguously signaling enhanced Kondo screening at lower temperatures. This screening mechanism effectively quenches magnetic fluctuations, thereby restoring single-particle coherence in the low-temperature regime —a behavior that stands in stark contrast to conventional magnetic systems but is fully consistent with our experimental observations.

To explain the absence of long-range magnetic order in $CsCr_6Sb_6$, we conducted DFT+$U$ ($U$ = 4 eV) calculations to probe its magnetic ground state. Our results reveal AFM coupling between the two kagome layers, with a preference for in-plane magnetism over out-of-plane alignment, consistent with prior studies. We systematically evaluated various candidate magnetic configurations (Figs. 5(g)-(n); two David stars denote the upper and lower kagome layers within the same double-layer unit). Among these, the configuration in Fig. 5(l) exhibits the lowest energy. However, the energy difference between the ground state and the next-lowest-lying configuration is merely sub-meV—a negligible scale that underscores the near-degeneracy of competing magnetic states. Conventional DFT calculations often fail to capture short-range magnetic correlations, but our findings suggest strong local magnetic coupling in $CsCr_6Sb_6$. The minimal energy differences between configurations imply that long-range order is destabilized by the interplay of nearly degenerate magnetic phases. Consequently, we conclude that $CsCr_6Sb_6$ hosts short-range magnetic correlations arising from localized moments embedded in a highly screened electronic environment.

**Conclusion**

Our comprehensive investigation of $CsCr_6Sb_6$ reveals a remarkable interplay of quantum phenomena that challenges conventional paradigms in strongly correlated systems. Through coordinated experimental and theoretical approaches, we establish three fundamental breakthroughs: First, the observation of pronounced Kondo scattering alongside heavy-fermion behavior in magnetotransport measurements unambiguously demonstrates strong electronic correlations in this *d*-electron kagome system. Second, ARPES reveals the unprecedented coexistence of an isolated, dispersionless flat band at $E_F$ with correlation-driven renormalization effects - creating an ideal yet previously unrealized platform for studying magnetic correlations in flat-band systems. Most strikingly, ZF-$\mu$SR measurements uncover the formation of short-range magnetic order, signaling a ground state where Kondo screening vigorously competes with magnetic correlations rather than achieving complete screening as in conventional Kondo systems.

Theoretical analysis provides the crucial microscopic understanding of this unique quantum state. We demonstrate that strong electronic correlations serve as the dual driver for both the dramatic mass renormalization of the flat band and the emergence of short-range magnetic order. This creates a quantum-critical landscape on the kagome lattice where frustrated magnetism and Kondo physics achieve delicate coexistence.

The contrast with CsCr$_3$Sb$_5$'s long-range AFM order is particularly instructive: while both systems feature flat bands, CsCr$_6$Sb$_6$ occupies a more nuanced regime near criticality where competition between interactions produces short-range order alongside heavy-fermion behavior.

These findings collectively establish CsCr$_6$Sb$_6$ as a prototypical system that: (1) challenges the *f*-electron hegemony in heavy-fermion physics through its *d*-electron correlated flat band; (2) provides the clearest realization to date of the Kondo-geometric frustration competition, with short-range order as its definitive experimental signature; and (3) reveals a pathway to novel quantum criticality through the interplay of Kondo and magnetic correlations. The emergence of such robust correlation effects in a non-*f*-electron system not only expands the horizons for discovering exotic quantum states but fundamentally reshapes our understanding of where and how strongly correlated physics can manifest. This work therefore opens multiple frontiers - from the controlled engineering of flat-band correlated materials to the exploration of new non-Fermi-liquid states arising from the rich interplay of Kondo physics, frustrated magnetism, and electronic correlations in quantum-critical systems.

**Acknowledgements**

The authors acknowledge the National Key R&D Program of China (Grants No. 2024YFA1408400, 2022YFA1402703, and 2023YFA140610) and the National Nature Science Foundation of China (Grants No. 92365204, 12374063 and 12174065). Y.F.G. acknowledges the open research fund of Beijing National Laboratory for Condensed Matter Physics (2023BNLCMPKF002). G.L. is supported by the Sino-German Mobility program (No. M-0006) and Shanghai 2021- Fundamental Research Area (No. 21JC1404700). Q.B.L. acknowledges support from the Ministry of Science and Technology of China (2023YFA1407400) and the Shanghai Natural Science Fund for Original Exploration Program (23ZR1479900). L.S. thanks the support by Innovation Program for Quantum Science and Technology (Grant No. 2024ZD0300104). The authors also thank the Analytical Instrumentation Center (#SPST-AIC10112914) and

# SI

## A. Crystal growth, characterizations and magnetotransport measurements

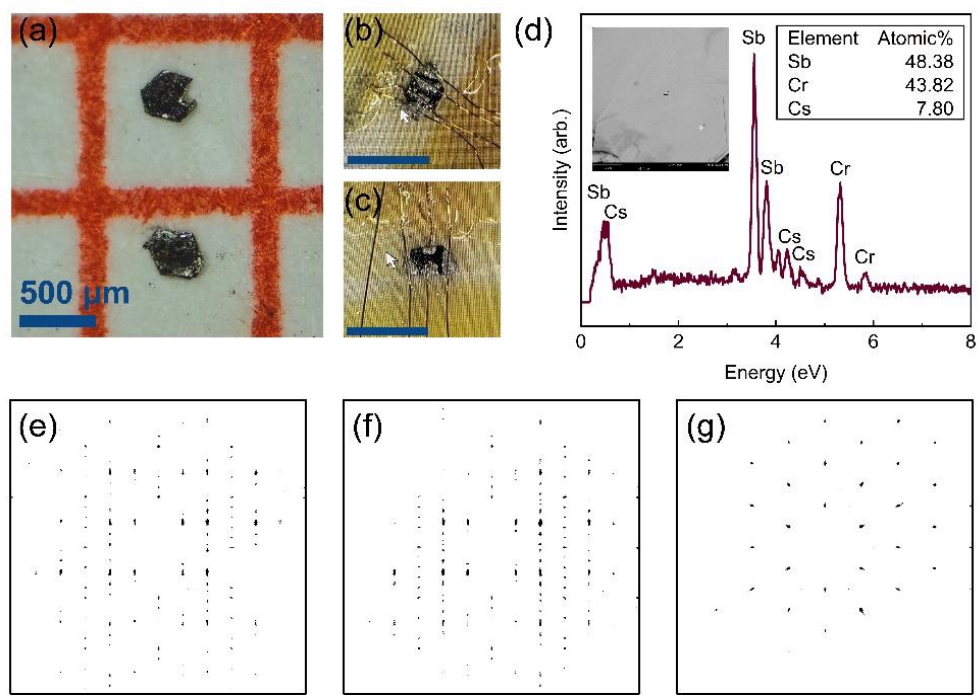

**Fig. S1.** (a) Picture of two typical CsCr$_6$Sb$_6$ single crystals. (b) and (c) are the optical image of sample S1 for MR and Hall measurements, respectively. (d) EDS spectrum of CsCr$_6$Sb$_6$. (e - g) SXRD patterns in reciprocal space on (0$kl$), ($h$0$l$) and ($hk$0) planes, respectively.

Single crystals of CsCr$_6$Sb$_6$ were synthesized via a self-flux method similar as that in Ref. [37]. High-purity starting materials - Cs liquid (99.98%), Cr powder (99.9%), and Sb granules (99.999%) - were combined in a 1:1:5 molar ratio, sealed in an alumina crucible within a tantalum tube under vacuum. The thermal protocol consisted of: (1) heating to 1100°C (15 h), (2) controlled cooling to 900°C (5 h), (3) isothermal annealing (10 h), and (4) slow cooling to 500°C (1.5°C/h), followed by air quenching. Residual flux was removed through selective etching of Cs-Sb phases using ultrapure water.

Crystal composition was verified by energy-dispersive X-ray spectroscopy (EDS)

using a Phenom Pro SEM. Structural characterization included: (i) single-crystal X-ray diffraction (Bruker D8, Mo K$_{\alpha1}$ radiation, λ = 0.71073 Å, 278 K) analyzed with Olex2 [38], and (ii) powder XRD (Bruker D8 Advance, Cu K$_{\alpha1}$ radiation, λ = 1.54184 Å, 298 K) refined using GSAS-II [39].

Transport properties were investigated in a Quantum Design DynaCool PPMS using standard Hall bar geometry for simultaneous resistivity and Hall effect measurements. Magnetic susceptibility was determined using a Quantum Design MPMS3 system.

## B. Magnetotransport data analysis

The interplay between magnetism and electronic structure is revealed through comprehensive magnetotransport measurements. Fig. S2(b) displays the temperature-dependent Hall resistivity $\rho_{xy}$ (1.8-250 K) for sample S1, showing pronounced nonlinearity at low temperatures. The distinct field-dependent behavior manifests as two linear regimes with different slopes ($R_H = d\rho_{xy}/dB$), whose bifurcation below $T \approx T_K$ is quantified in Fig. S3. The anomalous Hall contribution, obtained by extrapolating the high-field $\rho_{xy}$ to zero field [13, 40], is not discussed in detail here.

The observed field-dependent nonlinearity unequivocally demonstrates multi-carrier transport. We therefore employ a two-band model to extract carrier densities [41]:

$$\rho_{xy}(B) = 1/e \cdot \frac{(n_h\mu_h^2 - n_e\mu_e^2) + (n_h - n_e)\mu_e^2\mu_h^2 B^2}{(n_h\mu_h + n_e\mu_e)^2 + (n_h - n_e)\mu_e^2\mu_h^2 B^2} B, \qquad (1)$$

where $n_e(n_h)$ and $\mu_e(\mu_h)$ represent the electron (hole) density and mobility, respectively. The fitting results are presented in Table S1. Figs. S4(a,b) present excellent agreement between the model (dashed lines) and experimental data across both low- (2-20 K) and high-temperature (30-250 K) regimes.

**Table S1.** Longitudinal resistivity ($\rho_{xx}$), longitudinal conductivity ($\sigma_{xx}$), carrier concentration ($n_e$) and carrier mobility ($\mu_e$) of electrons, carrier concentration ($n_h$) and carrier mobility ($\mu_h$) of holes in $CsCr_6Sb_6$, recorded at 1.8 K and 40 K.

| $T$ | $\rho_{xx}$ | $\sigma_{xx}$ | $n_e$ | $\mu_e$ | $n_h$ | $\mu_h$ |
| --- | --- | --- | --- | --- | --- | --- |
| K | $\mu\Omega$ cm | $10^3$ $\Omega^{-1}$ cm$^{-1}$ | $10^{19}$ cm$^{-3}$ | cm$^2$ V$^{-1}$ s$^{-1}$ | $10^{19}$ cm$^{-3}$ | cm$^2$ V$^{-1}$ s$^{-1}$ |
| 1.8 | 184.48 | 5.42 | 3.19 | 115.25 | 1.95 | 577.21 |
| 40 | 184.48 | 5.42 | 8.95 | 377.21 | | |

Fig. S4(c) displays the temperature-dependent carrier density $n$ derived from fitting results. It is apparent that the carrier density $n$ undergoes a significant reduction as temperature decreases, a behavioral signature that strongly points to the onset of heavy fermion physics accompanied by an ultra-low carrier density. For a more refined perspective, the magnified inset in this figure further confirms the presence of a multi-carrier regime below the Kondo temperature $T_K$. Within this regime, the hole and electron densities tend to converge toward a near-compensated state, which is critical for unraveling the underlying electronic properties associated with band renormalization. The evolutionary trends of magnetism and electronic structure, as reflected in MR and Hall effect behaviors, are consistently reproducible across two additional distinct bulk samples (S2 and S3, Fig. S5), thereby reinforcing the generality of these phenomena.

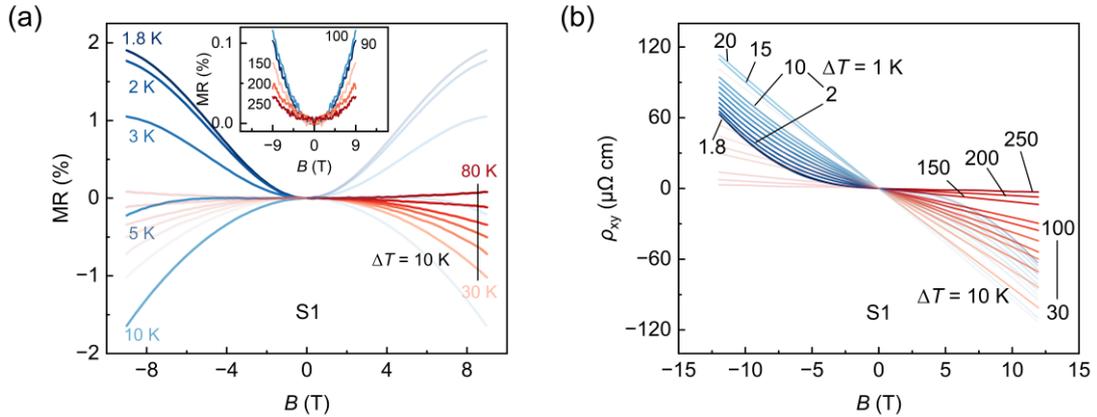

**Fig. S2.** (a) MR and (b) Hall resistivity at selected temperatures ($H//c$ up to 12 T), measured on the bulk sample in the main text. The inset in (a) shows the weak positive MR at high temperature.

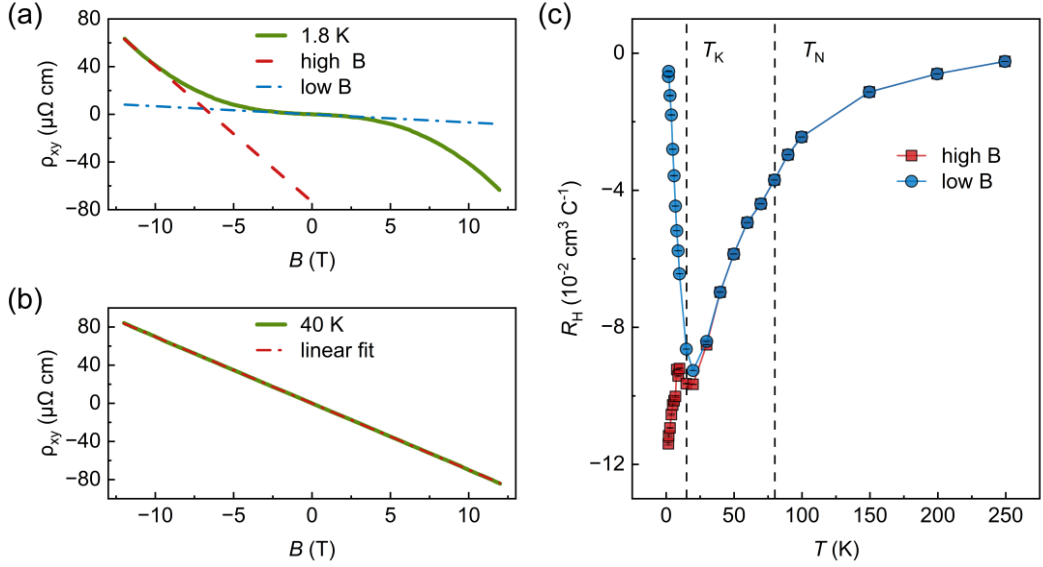

**Fig. S3.** (a, b) Linear fitting of $\rho_{xy}$ at representative temperatures. Below $T_K$, the anomalous Hall contribution is isolated by extrapolating the high-field linear regime to zero field. (c) Temperature dependence of $R_H$ extracted from both low- and high-field regions, revealing a clear bifurcation below $T_K$ that signifies the onset of competing electronic states.

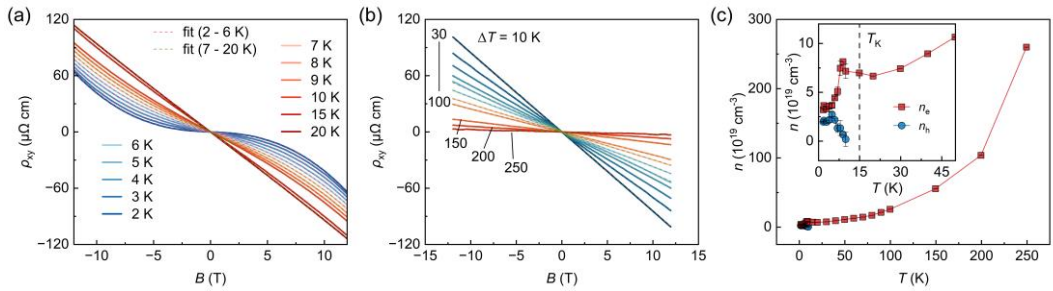

**Fig. S4.** (a, b) $\rho_{xy}$ of bulk sample S1 (main text), with dashed lines representing two-band model fits. While high-temperature data exhibit linear single-band transport, the emergent nonlinearity below $T_K$ necessitates a two-carrier description. (c) Temperature dependent carrier density $n$ derived from fit. The carrier density $n$ decreases rapidly with temperature, revealing the formation of heavy fermion behavior with ultra-low carrier density. The magnified inset further evidences a multi-

carrier regime below $T_K$, where hole and electron densities appear to approach near-compensation.

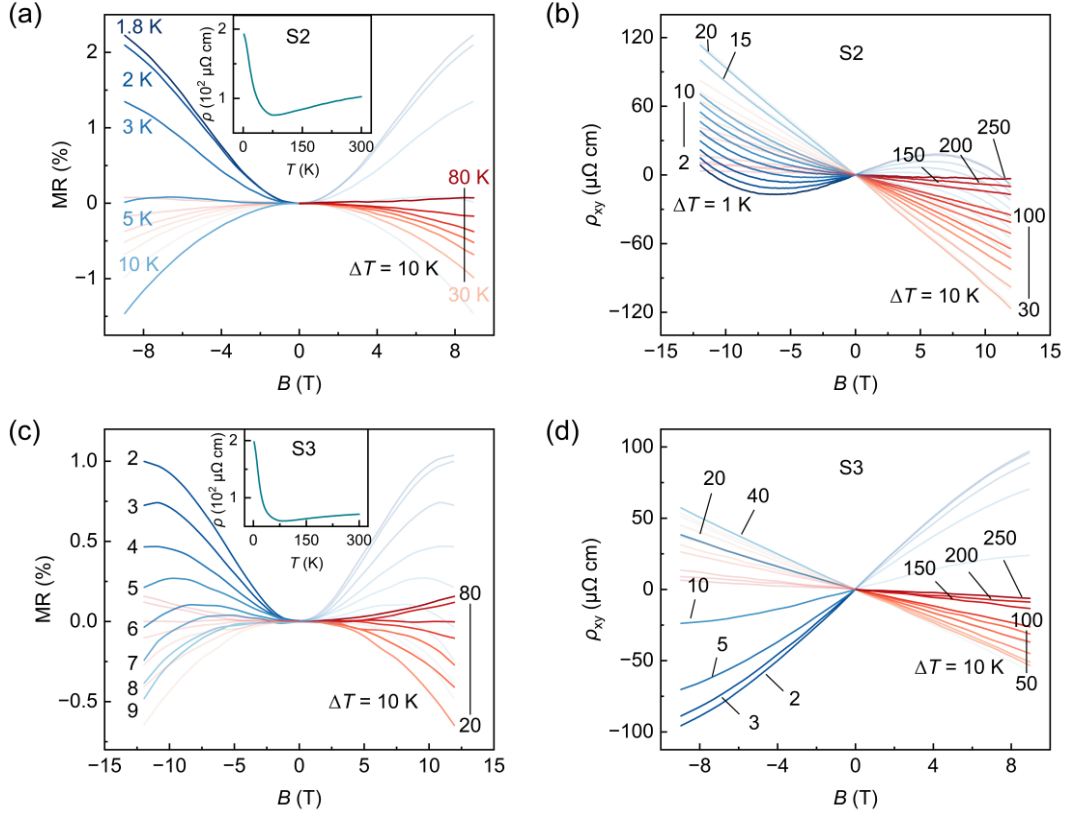

**Fig. S5.** (a, c) MR and (b, d) $\rho_{xy}$ measurements for bulk samples S2 and S3 under magnetic fields up to 12 T ($H//c$). The MR evolution traces the development of magnetic correlations, while the Hall response directly probes band renormalization effects. Notably, the insets in (a) and (c) reveal the characteristic $-\ln T$ dependence of longitudinal resistance, a hallmark of Kondo scattering.

## C. Angle-resolved photoemission spectroscopy (ARPES) measurements

High resolution ARPES measurements were performed at the "Dreamline" (BL09U) beamline and the BL03U beamline of the Shanghai Synchrotron Radiation Facility (SSRF) [42], using a SCIENTA Omicron DA30L analyzer. Samples were cleaved *in situ* under ultrahigh vacuum ($< 5\times10^{-11}$ Torr) and measured with energy and angular resolutions of 10–60 meV and 0.1°, respectively.

## D. Zero-field muon spin relaxation (*μ*SR) measurements

Zero-field *μ*SR experiments were conducted at TRIUMF's M20 and M9 beamlines (Vancouver, Canada) to detect local magnetic order. Crystals were mounted with the *c*-axis normal to a silver holder, and muon spins were implanted parallel to the *ab*-plane in spin-rotate (SR) mode. Data collected down to 2 K were analyzed using MUSRFIT to extract relaxation dynamics [43].

## E. First-principles calculations

First-principles calculations employed the Vienna *Ab Initio* Simulation Package (VASP) [44, 45] with the PBE+*U* functional ($U = 4.0$ eV, Dudarev scheme) and spin-orbit coupling [46]. A 400 eV plane-wave cutoff and 9×9×9 **k**-mesh ensured convergence. Correlation effects were treated via charge self-consistent DFT+DMFT using Wien2k (GGA-PBE, $RK_{max} = 7.0$, 10×10×10 **k**-mesh) and the embedded-DMFT package [47, 48]. The impurity problem for Cr-3*d* orbitals ($U = 2$–4 eV, $J = U/5$) was solved by continuous-time quantum Monte Carlo (CT-QMC) [49-52], with spectral functions computed via maximum entropy analytic continuation.

## F. Modulation of Kondo effect through pressure

High pressure serves as an ideal tuning parameter for electronic properties, enabling precise control of exchange interactions and electronic correlations without introducing chemical disorder. In CsCr$_6$Sb$_6$, the weak van der Waals bonding between layers facilitates significant structural modifications under pressure. We systematically investigate the pressure evolution of the Kondo lattice and heavy fermion behavior through combined resistivity and XRD measurements.

As shown in Fig. S6(a), the low-temperature resistivity upturn which is a signature of Kondo scattering, is progressively suppressed with increasing pressure, disappearing completely above 1.8 GPa. This pressure-induced Kondo breakdown is further evidenced by the vanishing -ln$T$ dependence, suggesting a fundamental change in the electronic ground state. The comprehensive phase diagram constructed from these measurements (Fig. S6(b)) reveals three distinct regimes: (1) a Kondo-dominated phase with $T_N$ decreasing from 80 K (ambient pressure) to 10 K (1.8 GPa), (2) a metallic phase persisting down to ~10 K at higher pressures, and (3) a weakly insulating regime emerging at the lowest temperatures, whose origin warrants further investigation.

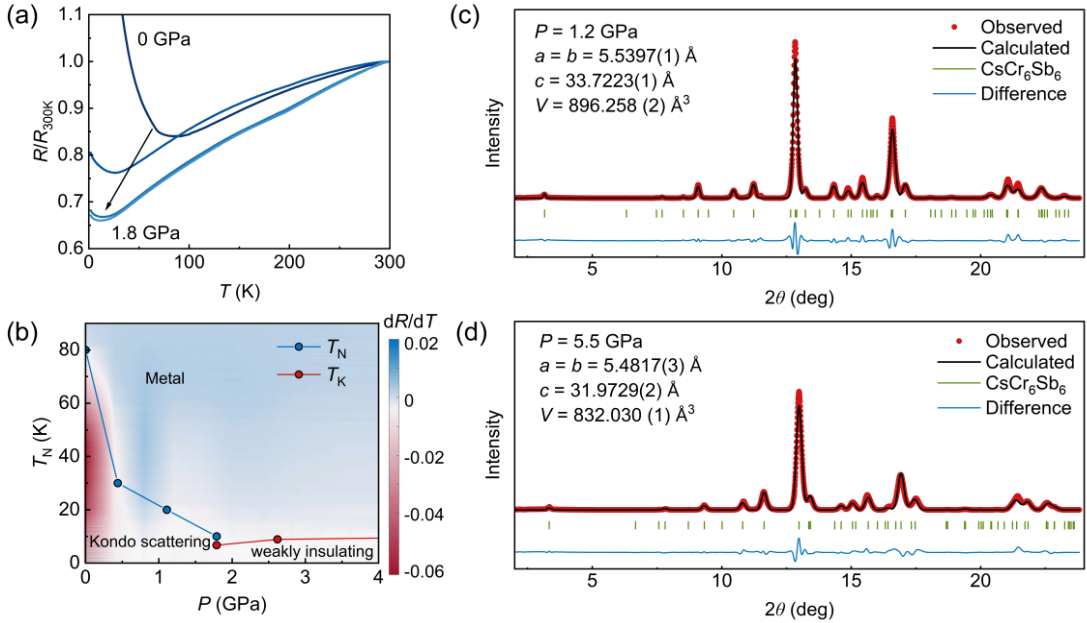

**Fig. S6.** Pressure-dependent electronic and structural evolution. (a) Temperature evolution of electrical resistance in CsCr$_6$Sb$_6$ under hydrostatic pressure (0-1.8 GPa). (b) Color-mapped d$R$/d$T$ plot revealing pressure dependence of the Néel temperature ($T_N$, determined from resistivity minima) and Kondo temperature ($T_K$, marking deviation from -ln$T$ behavior). Three distinct electronic regimes emerge: Kondo-dominated, metallic, and weakly insulating. (c,d) High-pressure structural integrity confirmed by Rietveld refinement of XRD patterns at 1.2 GPa and 5.5 GPa, respectively.

To elucidate the underlying mechanisms, we track the structural evolution through high-pressure XRD. Rietveld refinement of patterns collected up to 30.3 GPa (Fig. S7(a)) confirms the absence of structural phase transitions within the studied pressure range. The lattice compression exhibits pronounced anisotropy (Fig. S7(b)), with the *c*- axis (van der Waals gap) compressing more rapidly than the ab-plane—a direct consequence of the layered crystal architecture. Representative refinements at 1.2 GPa and 5.5 GPa (Figs. S6(c,d)) provide the structural parameters essential for modeling pressure-modified electronic states.

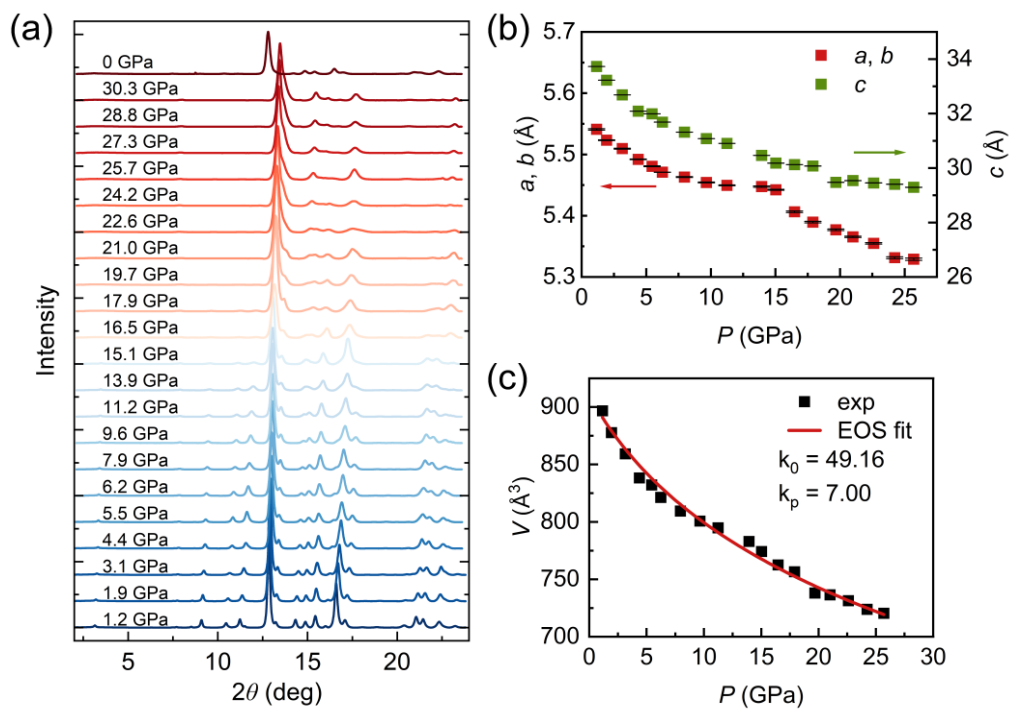

**Fig. S7.** (a) XRD patterns of $CsCr_6Sb_6$ under various pressures between 1.2 GPa and 30.3 GPa. Lattice parameters (b) and cell volume (c) as a function of the pressure. The red line in (c) is an equation of state (EOS) fit of cell volume under pressure.